\documentclass[aps,nofootinbib]{revtex4}
\usepackage[dvips]{graphicx}
\usepackage{latexsym}
\usepackage{revsymb}
\usepackage{amsfonts}
\usepackage{amsmath}
\usepackage{amssymb}
\usepackage{bm}

\newcommand{\ket}[1]{| #1 \rangle}

\newcommand{\bbra}[1]{\bigl\langle #1 \bigr|}
\newcommand{\bket}[1]{\bigl| #1 \bigr\rangle}
\newcommand{\bbraket}[2]{\bigl\langle #1 \big| #2\bigr\rangle}

\newcommand{\bvev}[1]{\bigl\langle #1 \bigr\rangle}

\newcommand{\dvev}[1]{\langle\!\langle #1 \rangle\!\rangle}
\newcommand{\bdvev}[1]{\bigl\langle\!\bigl\langle #1 \bigl\rangle\!\bigr\rangle}

\newcommand{\btvev}[1]{\bigl\langle\!\bigl\langle\!\bigl\langle #1 \bigl\rangle\!\bigl\rangle\!\bigr\rangle}

\newcommand{\Ahat}{\hat{A}}
\newcommand{\Bhat}{\hat{B}}
\newcommand{\Chat}{\hat{C}}

\begin{document}

\title{Bell's Inequalities, Superquantum Correlations, and String Theory}

\author{Lay Nam Chang}\email{laynam@vt.edu}
\author{Zachary Lewis}\email{zlewis@vt.edu}
\author{Djordje Minic}\email{dminic@vt.edu}
\author{Tatsu Takeuchi}\email{takeuchi@vt.edu}
\author{Chia-Hsiung Tze}\email{kahong@vt.edu}

\affiliation{Department of Physics, Virginia Tech, Blacksburg, VA 24061, USA}
\date{April 17, 2011}

\begin{abstract}
We offer an interpretation of super-quantum correlations in terms of
a ``doubly'' quantum theory. 
We argue that string theory, viewed as a quantum theory with two deformation
parameters, the string tension $\alpha'$ and
the string coupling constant $g_s$, is such a super-quantum theory, one
that transgresses the usual quantum violations of Bell's inequalities.
We also discuss the $\hbar \to \infty$ limit of quantum mechanics in this context.
As a super-quantum theory, string theory should display distinct
experimentally observable super-correlations
of entangled stringy states.
\end{abstract}

\maketitle
\section{Introduction}

In this note, we present an observation relating two fields of physics
which are ostensibly quite remote, namely: the study of the foundations of
quantum mechanics (QM) centered around the violation of the celebrated
Bell inequalities \cite{bell,CHSH}, and string theory (ST) \cite{string}.
As is well known, the Bell inequalities, based on the assumption of classical
local realism, are violated by the correlations of canonical QM \cite{exp}.
This remarkable feature of QM is often called ``quantum non-locality,''
though perhaps a misnomer.
However, even quantum correlations, with their apparent ``non-locality,'' are bounded 
and satisfy another inequality discovered by Cirel'son\footnote{Also spelled Tsirelson} \cite{cirelson}.
The natural question that arises is: do ``super''quantum theories exist which predict
correlations that transcend those of QM and thereby violate the Cirel'son bound?
Popescu and Rohrlich have demonstrated that such ``super''correlations 
can be consistent with relativistic causality (aka the no-signaling principle) \cite{super}. 
But what theory would predict them?
In the following, we give heuristic arguments which suggest that
non-perturbative ST may precisely be such a ``super-quantum'' theory.

\section{Bell's Inequality, the Cirel'son bound, and beyond}

Consider two classical variables $A$ and $B$, which represent the outcomes of 
measurements performed on some isolated physical system by detectors 1 and 2 placed
at two causally disconnected spacetime locations.
Assume that the only possible values of both $A$ and $B$ are $\pm 1$.
Denote the state of detector 1 by $a$, and that of detector 2 by $b$.
``Local realism'' demands that $A$ depend only on $a$, and $B$ depend only on $b$.
They can also depend on some hidden, but shared, information, $\lambda$. 
The correlation between $A(a,\lambda)$ and $B(b,\lambda)$ is then
\begin{equation}
P(a, b) \;=\; \int d\lambda\; \rho(\lambda)\, A(a,\lambda)\, B(b,\lambda)\;,
\qquad
\int d\lambda\,\rho(\lambda) \;=\; 1\;,
\label{Ccorr}
\end{equation}
where $\rho(\lambda)$ is the probability density of the hidden information $\lambda$.
This classical correlation is bounded by the following form of Bell's inequality \cite{bell}
as formulated by Clauser, Horne, Shimony and Holt (CHSH) \cite{CHSH}:
\begin{equation}
\Bigl|
 P(a, b) 
+P(a, b')
+P(a', b) 
-P(a', b') 
\Bigr| \;\leq\; X\;,
\qquad \mbox{where $X\;=\;X_\mathrm{Bell}\;=\;2\;.$}
\label{CHSHbound}
\end{equation}
The quantum versions of these correlations violate this bound, 
but are themselves bounded by a similar inequality obtained by replacing the
value of $X$ on the right-hand side with $X_\mathrm{QM}=2\sqrt{2}$. 
This is the famous Cirel'son bound \cite{cirelson},
the extra factor of $\sqrt{2}$ being determined by the
Hilbert space structure of QM.
The same Cirel'son bound has been shown to apply 
for quantum field theoretic (QFT) correlations also \cite{qft}.

Let us briefly review the simplest routes to these bounds.
Following Refs.~\cite{cirelson} and \cite{dieks},
consider 4 classical stochastic variables $A$, $A'$, $B$, and $B'$, each of which
takes values of $+1$ or $-1$. 
Obviously, the quantity 
\begin{equation}
C \;\equiv\; AB+AB'+A'B-A'B' \;=\; A(B+B')+A'(B-B')\;,
\end{equation}
can be only $+2$ or $-2$
and thus the absolute value of its expectation value is bounded by 2:
\begin{equation}
\Big|\bvev{C}\Bigr| 
\;=\;
\Bigl|\bigl\langle AB+ AB'+A'B-A'B' \bigr\rangle\Bigr| 
\;\leq\; 2\;.
\end{equation}
This is the classical Bell bound.
For the quantum case, we replace the classical stochastic variables 
with hermitian operators acting on a Hilbert space.
Following Ref.~\cite{cirelson}, we find that
if $\Ahat^2=\Ahat'{}^2=\Bhat^2=\Bhat'{}^2=1$ and
$\bigl[\Ahat , \Bhat\bigr]=
\bigl[\Ahat , \Bhat'\bigr]=
\bigl[\Ahat' , \Bhat\bigr]=
\bigl[\Ahat' , \Bhat'\bigr]=0\;$,
then $C$ is replaced by
\begin{equation}
\Chat \;=\; \Ahat\Bhat + \Ahat\Bhat' + \Ahat'\Bhat - \Ahat'\Bhat'\;,
\end{equation}
from which we find
\begin{equation}
\Chat^2 \;=\; 4 - \bigl[\Ahat , \Ahat'\bigr]\!\cdot\!\bigl[\Bhat , \Bhat'\bigr]\;.
\end{equation}
When the commutators are zero,
we recover the classical bound of 2.
If they are not, we can use the uncertainty relations
$\bigl|\bvev{i\bigl[\Ahat , \Ahat'\bigr]}\bigr|\le 2 \|\Ahat\|\!\cdot\!\|\Ahat'\|$ and
$\bigl|\bvev{i\bigl[\Bhat , \Bhat'\bigr]}\bigr|\le 2 \|\Bhat\|\!\cdot\!\|\Bhat'\|$ to obtain
\begin{equation}
\bvev{\Chat^2} 
\;\leq\; 4 + 4\; \|\Ahat\|\!\cdot\!\|\Ahat'\|\!\cdot\!\|\Bhat\|\!\cdot\!\|\Bhat'\| 
\;=\; 8 
\qquad\longrightarrow\qquad 
\Bigl|\bvev{\Chat}\Bigr| 
\;\leq\; \sqrt{\bvev{\Chat^2}}
\;\leq\; 2\sqrt{2}\;,
\end{equation}
which is the Cirel'son bound.
Alternatively, we can follow Ref.~\cite{dieks} and let 
$\Ahat\ket{\psi} = \ket{A}$,
$\Bhat\ket{\psi} = \ket{B}$, 
$\Ahat'\ket{\psi} = \ket{A'}$, and 
$\Bhat'\ket{\psi} = \ket{B'}$.
These 4 vectors all have unit norms and  
\begin{equation}
\left|\bvev{\Chat}\right|  
\;=\; \left|\bbra{\psi}\Chat\bket{\psi}\right|
\;=\; \Bigl|\bbraket{A}{B+B'} +\bbraket{A'}{B-B'} \Bigr|
\;\leq\;
\Bigl\| \,\ket{B}+\ket{B'}\, \Bigr\| + 
\Bigl\| \,\ket{B}-\ket{B'}\, \Bigr\|
\;,
\end{equation}
which implies:
\begin{equation}
\left|\bvev{\Chat}\right| 
\;\leq\;
\sqrt{2\left(1+ \mathrm{Re}\,\bbraket{B}{B'}\right)} +
\sqrt{2\left(1- \mathrm{Re}\,\bbraket{B}{B'}\right)} 
\;\leq\; 2\sqrt{2}
\;.
\end{equation}
This second proof suggests that the Cirel'son bound is actually independent of the 
requirement of relativistic causality. If relativistic causality is broken, then
the $\Ahat$'s and $\Bhat$'s will not commute. Then $\Chat$ must be symmetrized as
\begin{equation}
\Chat \;=\; \dfrac{1}{2}
\left[
\left(\Ahat\Bhat+\Bhat\Ahat\right)
+\left(\Ahat\Bhat'+\Bhat'\Ahat\right)
+\left(\Ahat'\Bhat+\Bhat\Ahat'\right)
-\left(\Ahat'\Bhat'+\Bhat'\Ahat'\right)
\right]\;,
\end{equation}
to make it hermitian, and its expectation value will be
\begin{equation}
\bvev{\Chat} \;=\;
\mathrm{Re}\Bigl[\bbraket{A}{B+B'} +\bbraket{A'}{B-B'}\Bigr]
\;,
\end{equation}
which is clearly subject to the same bound as before.
So it is the Hilbert space structure of QM alone which determines this bound.
%


Indeed, Popescu and Rohrlich have demonstrated that one can
concoct super-quantum correlations which violate the Cirel'son bound,
while still maintaining consistency with relativistic causality \cite{super}.
However, such super-quantum correlations are also bounded, the value of $X$ in 
Eq.~(\ref{CHSHbound}) being replaced, not by $X_\mathrm{QM}=2\sqrt{2}$ but, by $X=4$:
\begin{equation}
\Bigl|
 P(a, b)
+P(a, b')
+P(a', b) 
-P(a', b') 
\Bigr| \;\leq\; 4 \;.
\label{FF}
\end{equation}
Note, though, that this is not a `bound' per se,
the value of 4 being the absolute maximum that the left-hand side can possibly be, 
since each of the $4$ terms has its absolute value bounded by one.
If the four correlations represented by
these $4$ terms were completely independent, then,
in principle, there seems to be no reason why this bound cannot be saturated.

But what type of theory would predict such correlations?
It has been speculated that a specific super-quantum theory 
could essentially be derived from the two requirements 
of relativistic causality and the saturation of the $X=4$ bound,
in effect elevating these requirements to the status of `axioms' 
which define the theory \cite{super}.
In a similar fashion, QM may also be derivable from causality and the Cirel'son bound 
as `axioms' \cite{aharonov}.
However, to our knowledge, no concrete realization of such a program has thus far emerged.

A related development has been the proof by van Dam that
super-quantum correlations which saturate the $X=4$ bound can be used to render 
all communication complexity problems trivial \cite{vanDam}. 
Subsequently, Brassard et al. discovered a protocol 
utilizing correlations with $X>X_\mathrm{cc}=4\sqrt{2/3}$, which 
solves communication complexity problems trivially in a probabilistic manner \cite{brassard}. 
Due to this, it has been speculated that nature somehow 
disfavors super-quantum theories, and that super-quantum correlations,
especially those with $X>X_\mathrm{cc}$, should not exist \cite{trivial}.  
However, the argument obviously does not preclude the existence of super-quantum theories
itself.

One proposal for a super-quantum theory discussed in the literature uses a formal mathematical 
redefinition of the norms of vectors from the usual $\ell^2$ norm to the more general 
$\ell^p$ norm \cite{pnorm}.  
In a 2D vector space with basis vectors $\{\mathbf{e}_1,\mathbf{e}_2\}$,
the $\ell^p$ norm is
\begin{equation}
\Bigl\|\,\alpha \mathbf{e}_1+\beta\mathbf{e}_2 \,\Bigr\|_p \;=\;
\sqrt[p]{\alpha^p+\beta^p}\;.
\end{equation}
If one identifies $\ket{B}=\mathbf{e}_1$ and $\ket{B'}=\mathbf{e}_2$, then
\begin{equation}
\Bigl\| \,\ket{B}\pm\ket{B'}\, \Bigr\|_p
\;=\; 2^{1/p}\;.
\label{pNorm}
\end{equation}
Eq.~(\ref{FF}) would then be saturated for the $p = 1$ case.\footnote{%
The $\ell^1$ norm and $\ell^\infty$ norm are equivalent in 2D, requiring a mere $45{}^\circ$
rotation of the coordinate axes to get from one to the other.
}
Unfortunately, it is unclear how one can construct a physical theory based on this proposal
in which dynamical variables evolve in time while preserving total probability.

At this point, we make the very simple observation that 
it is the procedure of ``quantization,'' which takes us from classical mechanics to QM,
that increases the bound from the Bell/CHSH value of 2 to the Cirel'son value of $2\sqrt{2}$. 
That is, ``quantization'' increases the bound by a factor of $\sqrt{2}$.
Thus, if one could perform another step of ``quantization'' onto QM, would it
not lead to the increase of the bound by another factor of $\sqrt{2}$, 
thereby take us from the Cirel'son value of $2\sqrt{2}$ to the ultimate $4$?
This is the main conjecture of this letter, that is,
a ``doubly'' quantized theory would lead to the violation of the Cirel'son bound.

In the following, we will clarify which  
``quantization'' procedure we have in mind, and how it can be applied for a second time onto QM,
leading to a ``doubly quantized'' theory.
We then argue that a physical realization of such a theory may be offered by 
non-perturbative open string field theory (OSFT).

\section{``Double'' Quantization and Open String Field Theory}

Before going into the ``double quantization'' procedure,
let us first observe that, 
from the point of view of general mathematical deformation theory \cite{fadd},
QM is a theory with one deformation parameter $\hbar$, while ST is a theory with two:
The first deformation parameter of ST is the world-sheet coupling constant
$\alpha'$, which measures the essential non-locality of the string, and is responsible
for the organization of perturbative ST.
The second deformation parameter of ST is the string coupling constant $g_s$, 
which controls the non-perturbative aspects of ST,
such as D-branes and related membrane-like solitonic excitations,
and the general non-perturbative string field theory (SFT) \cite{string}.
Therefore, ST can be expected to be more ``quantum'' in some sense 
than canonical QM, given the presence of the second deformation parameter.

Second, super-quantum correlations point to a non-locality, which is more
non-local, so to speak, than the aforementioned ``quantum non-locality'' of QM and QFT.
However, QFT's are actually local theories, and 
true non-locality is expected only in theories of quantum gravity.
That quantum gravity must be non-local stems from the requirement of diffeomorphism invariance,
as has been known from the pioneering days of that field \cite{bryce}.
Thus quantum gravity, for which ST is a concrete example,
can naturally be expected to lead to correlations more non-local than those in QM/QFT.

Third, the web of dualities discovered in ST \cite{string}, 
which points to the unification of QFT's in various dimensions,
can themselves be considered a type of ``correlation'' which transcends the barriers of
QFT Lagrangians and spacetime dimensions.
Again, the evidence suggests ``super'' correlations, perhaps much more ``super'' than
envisioned above, in the context of ST.

What follows is a heuristic attempt to make these expectations physically concrete.
Our essential observation is as follows:
The ``quantization'' procedure responsible for 
turning the classical Bell bound of $2$ into 
the quantum Cirel'son bound of $2\sqrt{2}$ is
given by the path integral over the classical dynamical variables, 
which we collectively denote as $x$.
That is,
given a classical action $S(x)$,
functions of $x$ are replaced by their expectation values defined via the path integral
\begin{equation}
f(x) \quad\rightarrow\quad 
\bvev{\,f(\hat{x})\,}  
\;=\; \int \!Dx\;f(x)\,\exp\left[\frac{i}{\hbar} S(x)\right]
\;,
\end{equation}
up to a normalization constant. 
In particular, the correlation between two observables
$\Ahat(a)$ and $\Bhat(b)$ will be given by
\begin{equation}
\bvev{\,\Ahat(a)\Bhat(b)\,}  \;=\; \int \!Dx\;A(a,x)\,B(b,x)\,\exp\left[\dfrac{i}{\hbar} S(x)\right]
\;\equiv\;
A(a)\star B(b)
\;.
\label{AstarB}
\end{equation}
\textit{cf}. Eq.~(\ref{Ccorr}).
In a similar fashion, we can envision taking a collection of quantum operators,
which we will collectively denote by $\hat{\phi}$, 
for which a ``quantum'' action $\tilde{S}(\hat{\phi})$ is given, and
define another path integral over the quantum operators $\hat{\phi}$ :
\begin{equation}
F(\hat{\phi}) \quad\rightarrow\quad 
\bdvev{\,F(\hat{\hat{\phi}}\,)}  
\;=\; \int \!D\hat{\phi}\;F(\hat{\phi})\,\exp\left[\frac{i}{\tilde\hbar} \tilde{S}(\hat{\phi})\right]
\;,
\end{equation}
and the correlation between two ``super'' observables will be
\begin{equation}
\bdvev{\,\hat\Ahat(a)\hat\Bhat(b)\,}  
\;=\; \int \!D\hat{\phi}\;\Ahat(a,\hat{\phi})\,\Bhat(b,\hat{\phi})\,\exp\left[\dfrac{i}{\tilde{\hbar}} \tilde{S}(\hat{\phi})\right]\;.
\end{equation}
Note that the expectation values here, denoted $\dvev{*}$, 
are not numbers but operators themselves.
To further reduce it to a number, we must calculate its expectation value in the usual way
\begin{equation}
\bdvev{\,\hat\Ahat(a)\hat\Bhat(b)\,}  
\quad\rightarrow\quad
\btvev{\,\hat\Ahat(a)\hat\Bhat(b)\,}  
\;=\;
\left\langle
\int \!D\hat{\phi}\;\Ahat(a,\hat{\phi})\,\Bhat(b,\hat{\phi})\,\exp\left[\dfrac{i}{\tilde{\hbar}} \tilde{S}(\hat{\phi})\right]
\right\rangle
\;,
\end{equation}
which would amount to replacing all the products of operators on the right-hand side
with their first-quantized expectation values, or equivalently, 
replacing the operators with `classical' variables except with their products
defined via Eq.~(\ref{AstarB}).

This defines our ``double quantization'' procedure, through which two 
deformation parameters, $\hbar$ and $\tilde{\hbar}$, are introduced.
We would like to emphasize that the $\hat{\phi}$ in the above expressions is already a quantum entity,
depending on the first deformation parameter $\hbar$. 
Thus the ``double quantization'' procedure proposed here is quite distinct from the
``second quantization'' procedure used in QFT, which, being a single quantization procedure 
of a classical field, is a misnomer to begin with.
The caveats to our definition are, of course, the difficulty in precisely defining the
path integral over the quantum operator $\hat{\phi}$, and thus doing any actual calculations with 
it, and imposing a physical interpretation on what is meant by the quantum operators 
themselves being probabilistically determined.

At this point, we make the observation that a ``doubly quantized'' theory may already be
available in the form of Witten's open string field theory (OSFT) \cite{witten}.
Our ``double'' quantization procedure can be mapped onto ST as follows:
In the first step, the classical action $S(x)$ can be identified with the world-sheet Polyakov action, 
and the first deformation parameter $\hbar$ with the world-sheet coupling $\alpha'$ \cite{string}.
In the second step, the quantum action $\tilde{S}(\hat{\phi})$ can be identified with 
Witten's OSFT action \cite{witten}, and the second deformation parameter $\tilde{\hbar}$ 
with the string coupling $g_s$.

The doubly deformed nature of the theory is
explicit in the Witten action for the `classical' open string field $\Phi$, 
an action of an abstract Chern-Simons type
\begin{equation}
S_W (\Phi) \;=\; \int \Phi \star Q_\mathrm{BRST} \Phi + \Phi \star \Phi \star \Phi\;,
\end{equation}
where $Q_\mathrm{BRST}$ is the open string theory BRST cohomology 
operator ($Q_\mathrm{BRST}^2=0$), and the star product is
determined via the world-sheet Polyakov action
\begin{equation}
S_P (X) \;=\; 
\dfrac{1}{2} \int \!d^2 \sigma\; \sqrt{\gamma}\, g^{ab}\, \partial_a X^i\, \partial_b X^j\, G_{ij} 
+ \cdots
\end{equation} 
and the corresponding world-sheet path integral
\begin{equation}
F \star G \;=\; \int \!D X\; F(X)\,G(X)\,\exp\left[\frac{i}{\alpha'} S_P (X)\right]\;.
\end{equation}
The fully quantum OSFT is then in principle defined by yet another
path integral in the infinite dimensional space of the open string field $\Phi$, i.e. 
\begin{equation}
\int D \Phi\,\exp\left[\dfrac{i}{g_s}S_W(\Phi)\right]\;,
\end{equation} 
with all products defined via the star-product.

In addition to its manifestly ``doubly'' quantized path integral, 
OSFT has as massless modes the ordinary photons, which are 
used in the experimental verification of the violation of
Bell's inequalities \cite{exp},
and it also contains gravity (closed strings) as demanded by
unitarity.\footnote{The open/closed string theory duality
is nicely illustrated by the AdS/CFT duality \cite{string}. It is interesting to contemplate
the Bell bound and its violations, both quantum and super-quantum, in this
well-defined context. Similarly, it would be interesting, even though experimentally
prohibitive, to contemplate the super-quantum correlations for the QCD string, perhaps
in the studies of the quark-gluon plasma.}
Thus, our heuristic reasoning suggests that OSFT may precisely
be an example of a super-quantum theory, which violates the Cirel'son bound.

We close this section with a caveat and a speculation.
In the above reasoning, the two quantizations were taken to be independent 
with two independent deformation parameters.
In the case of OSFT, they were $\alpha'$ and $g_s$.
However, from the point of view of M-theory, 
both these parameters would be determined dynamically and cannot be, in fact, independent.
Thus, it may not be correct to view OSFT as a fully ``doubly quantized'' theory.
Would this mean that OSFT/M-theory correlations would not saturate the ultimate $X=4$ bound?
Would its CHSH bound be situated somewhere between $X_\mathrm{QM}=2\sqrt{2}$
and $X=4$, perhaps below the communication complexity bound of $X_\mathrm{cc}=4\sqrt{2/3}$?

\section{The $\hbar\rightarrow\infty$ Limit}

Given that a super-quantum theory is supposedly more ``quantum'' than QM,
let us now consider the the extreme quantum limit of QM, 
$\hbar\rightarrow\infty$.
Though QM is not ``doubly quantized,'' could it still exhibit 
certain super-quantum behavior in that limit?
Taking a deformation parameter to infinity can be naturally performed in ST, 
either $\alpha'\rightarrow\infty$ or $g_s\rightarrow\infty$, 
and one can still retain sensible physics.
Therefore, the $\hbar\rightarrow\infty$ limit of QM may also be
a sensible theory, but at the same time quite different from QM.
After all, if the $\hbar\rightarrow 0$ limit is to recover classical mechanics,
with the Bell bound of $X_\mathrm{Bell}=2$, and apparently quite different from QM, 
it may not be too farfetched to conjecture that the 
$\hbar\rightarrow\infty$ limit would flow to a super-quantum theory, 
with the super-quantum bound of $X=4$.
If this were indeed the case, it may provide us with an opportunity to explore
super-quantum behavior in the absence of a solution to OSFT/M-theory.

What would the $\hbar\rightarrow\infty$ limit mean from the point of view
of the path integral?  Given that the path-integral measure is $e^{iS/\hbar}$,
in the $\hbar\rightarrow\infty$ limit this measure will be unity for any $S$,
and all histories in the path integral contribute with equal unit weight.
Similarly all phases, measured by $e^{iS/\hbar}$, 
will be washed out.\footnote{This immediately raises other issues, such as the meaning of quantum statistics.}
Because the phases are washed out, we cannot distinguish between
$\ket{B}+\ket{B'}$ and $\ket{B}-\ket{B'}$.\footnote{note that $-1 = e^{i \pi}$ and
that sign can be absorbed into a phase of $\ket{B'}$)} 
This suggests
\begin{equation}
\Bigl\|\,\ket{B} \pm \ket{B'}\,\Bigr\| 
\;=\; 
\Bigl\|\ket{B}\Bigr\| + \Bigl\|\,\ket{B'}\,\Bigr\|\;,
\label{BB}
\end{equation}
which, if applied to the proof of the Cirel'son bound given earlier, 
leads to the super-quantum bound of 4.
This property is similar to what was obtain by replacing the
$\ell^2$ norm with an $\ell^1$ (or $\ell^\infty$) norm, \textit{cf}. Eq.~(\ref{pNorm}), 
but presumably, unlike the change of norm, 
this relation is independent of the choice of basis.
This argument seems to suggest that the $\hbar\rightarrow\infty$ limit is indeed super-quantum.

However, this observation is perhaps a bit naive
since the proof of the Cirel'son bound itself may no longer be valid
under the washed-out of all phases.
Let us invoke here an optical-mechanical analogy: 
geometric optics is the zero wavelength limit of electromagnetism, 
which would correspond to the $\hbar\rightarrow 0$ limit of QM.
The $\hbar\rightarrow\infty$ limit of QM would therefore correspond 
the extreme near field limit of electromagnetism, and in that case,
the superposition of waves is washed out.\footnote{We thank Jean Heremans for discussions of this point.}
Note also, that from a geometric point of view, 
the holomorphic sectional curvature $2/\hbar$ of the projective Hilbert space $CP^N$ of canonical QM goes to zero as $\hbar\rightarrow\infty$, and $CP^N$ becomes just $C^N$.
(For a general discussion of the geometry of quantum theory and its relevance for quantum gravity and string theory, see \cite{chia}.)
From these observations, it is clear that the usual Born rule to obtain probabilities 
will no longer apply.

But before we ask what rule should replace that of Born, let us confront the 
obvious problem that in the limit $\hbar\rightarrow\infty$, only the ground state
of the Hamiltonian will remain in the physical spectrum and the theory will be rendered trivial.\footnote{%
If the system has a non-trivial topology, it could allow for degeneracies in the ground state, and thus lead to a non-trivial theory even in the $\hbar\rightarrow\infty$ limit.}
This can also be argued via the general Feynman-Schwinger formulation of QM \cite{feynman}:
\begin{equation}
\delta S\, \psi \;=\; i \hbar \,\delta \psi\;.
\end{equation}
By taking the $\hbar \to \infty$ limit, we eliminate
the classical part $\delta S$, so that we are left only with
$
\delta \psi = 0
$
and thus $\psi$ must be a constant $\psi \equiv |\psi|$, a trivial result.

Could the $\hbar\rightarrow\infty$ limit of QM be made less-than-trivial?
Consider the corresponding $\alpha'\rightarrow\infty$ and $g_s\rightarrow\infty$
limits in ST.
In the $\alpha' \to \infty$ of ST, as opposed to the
usual $\alpha' \to 0$ field theory limit,
one seemingly ends up with an infinite number
of fields and a non-trivial higher spin theory \cite{hs}.
Recently, such a theory was considered from a holographically
dual point of view, and the dual of such
a higher spin theory in AdS space was identified to
be a free field theory \cite{douglas}.
The $g_s\rightarrow\infty$ limit of ST appears in the context of M-theory, 
one of whose avatars arises in the $g_s \to \infty$ limit of type-IIA ST \cite{string}.
Neither the high spin theory, nor the avatars of M-theory are trivial,
the presence of the tunable second deformation parameter saving them from triviality.
Thus, the introduction of a second tunable parameter into QM, e.g.
Newton's gravitational constant $G_N$, may be necessary for the limit $\hbar\rightarrow\infty$
to be non-trivial.

Another issue here is that of interpretation:
in the classical ($\hbar \to 0$) case we have one trajectory, and one event
(position, for example) at one point in time.
One could speculate that the super-quantum ($\hbar \to \infty$) limit would correspond to the complement of
all other virtual trajectories. A general linear
map relating virtual and classical
trajectories is presumably non-symmetric (there are in principle more possibilities than
actual events).
Very naively, one would then expect that if we impose the condition that all possible events can be ``mapped'' to
actual events, we could end up with a symmetric linear map
corresponding to quantum theory ($\hbar \sim 1$), with
a natural ``map'' 
between the actual events and possibilities, presumably
realized by the Born probability rule.
Note that according to this scenario, the super-quantum theory
would correspond essentially to a theory of possibilities
and without actual events, which would be an interesting
lesson for the foundations of ST.

\section{Possible Experimental Signatures}

Finally we offer some comments on
possible experimental observations of such super-quantum violations of
Bell's inequalities.
The usual set-up involves entangled photons \cite{exp}.
In open ST, photons are  the lowest lying massless
states, but there is also an entire Regge trajectory associated with them.
So, the obvious experimental suggestion would be: observe entangled Reggeized photons.
Such an experiment is, of course, forbidding at present, given its Planckian nature.

Super-quantum correlations could also be observable in cosmology.
The current understanding of the large scale structure of the universe,
\textit{i.e.} the distribution of galaxies and galaxy clusters, 
is that they are seeded by quantum fluctuations.
In standard calculations, it is assumed that the quantum correlations
of these fluctuations are Gaussian.\footnote{Non-Gaussian correlations have also been considered.}
If the correlations were in fact super-quantum, however, 
their signature could appear as characteristic deviations from the
predicted large scale structure based on Gaussian correlations.

We conclude with a few words regarding a new experimental ``knob'' needed to test
our doubly quantized approach to super-quantum correlations.
In the classic experimental tests of the violation of Bell's inequalities \cite{exp}, such a ``knob'' is
represented by the relative angle between polarization vectors
of entangled photos.
If we have another
quantization, there should be, in principle, another angle-like ``knob.''
Thus, the usual one dimensional data plot \cite{exp} should be replaced by
a two dimensional surface. 
By cutting this surface at various values of the new, second angle, we should be able to
obtain one dimensional cuts for which the value of the CHSH bound varies depending on the cut: 
exceeding $2 \sqrt{2}$ in some cases, and perhaps not exceeding $2$ in others.
Thus, the second ``knob'' may very well allow us to interpolate between the classical,
quantum, and superquantum cases.

In this note, 
we have obviously only scratched the surface of a possible super-quantum theory, and many
probing questions remain to be answered and understood. 
We hope to address some of them in future works.

\section*{Acknowledgments} 

We thank V. Balasubramanian, J. Heremans, S. Mathur, K. Park, R. Raghavan, V. Scarola, 
and A. Staples for interesting discussions and questions.
DM acknowledges the hospitality of the Mathematics Institute at Oxford University
and Merton College, Oxford, and his respective hosts, Philip Candelas and Yang-Hui He. DM also thanks the Galileo Galilei Institute for Theoretical Physics, Florence, for the hospitality and the INFN for partial support.
ZL, DM, and TT are supported in part by
the U.S. Department of Energy, grant DE-FG05-92ER40677, task A.


\end{document}